# Laplacian Flow Dynamics on Geometric Graphs for Anatomical Modeling of Cerebrovascular Networks

Rafat Damseh[1], *Student Member, IEEE*, Patrick Delafontaine-Martel [1], Philippe Pouliot[2,3], Farida Cheriet[4], *Senior Member, IEEE,* and Frederic Lesage[1,2,3]

*Abstract*—Generating computational anatomical models of cerebrovascular networks is vital for improving clinical practice and understanding brain oxygen transport. This is achieved by extracting graph-based representations based on pre-mapping of vascular structures. Recent graphing methods can provide smooth vessels trajectories and well-connected vascular topology. However, they require water-tight surface meshes as inputs. Furthermore, adding vessels radii information on their graph compartments restricts their alignment along vascular centerlines. Here, we propose a novel graphing scheme that works with relaxed input requirements and intrinsically captures vessel radii information. The proposed approach is based on deforming geometric graphs constructed within vascular boundaries. Under a laplacian optimization framework, we assign affinity weights on the initial geometry that drives its iterative contraction toward vessels centerlines. We present a mechanism to decimate graph structure at each run and a convergence criterion to stop the process. A refinement technique is then introduced to obtain final vascular models. Our implementation is available on *https://github.com/Damseh/VascularGraph*. We benchmarked our results with that obtained using other efficient and state-of-the-art graphing schemes, validating on both synthetic and real angiograms acquired with different imaging modalities. The experiments indicate that the proposed scheme produces the lowest geometric and topological error rates on various angiograms. Furthermore, it surpasses other techniques in providing representative models that capture all anatomical aspects of vascular structures.

## I. INTRODUCTION

Modeling of cerebrovascular structures is essential in areas ranging from clinical decision support to fundamental research. Topological and structural representations of cerebrovascular networks substantially facilitate guided surgical procedures involving intracranial electrode placement [1], catheter motion planning [2] and endovascular aneurysm repair [3]. Computational modeling of cerebrovascular networks can also allow for better intersubject assessment of vascular features [4]. Beyond macro-scaled clinical images, studying cerebral microvasculature is vital for understanding brain neurovascular coupling and neuro-metabolic activity [5]. It is also important to investigate neuropathologies associated with a deterioration in cerebral oxygen transport [6], [7].

Clinically, non-invasive cerebrovascular imaging techniques including magnetic resonance angiography (MRA) and computed-tomography angiography (CTA) are common practices for both preoperative planning and postoperative surveillance scanning. In experimental studies, optical imaging systems have been recently proposed to provide spatially-resolved measurements of cerebral microvasculature in-vivo. [8]. Transforming the cerebrovascular structure represented within a quantized spatial grid into an interpretable computational model is problematic. Cerebrovascular images exhibit a high level of intersubject heterogeneity, and contain complex vascular structures. Furthermore, vascular space is submitted to dynamically evolving conditions like acute ischemic strokes. These obstacles hinder the construction of accurate computational models that encode connectivity and spatial information to be used for further analysis of shape features and hemodynamics properties.

## II. RELATED WORK

Automatic graph extraction schemes have generated interest in many fields spanning from urban planning to neuroinformatics [9]–[25]. For the use of topological and geometrical features in medical imaging applications, the reader is referred to [25]. Some graphing techniques [14]–[16] aim to reconstruct tree-like structures, whereas others target the modeling of loopy curvilinear structures [19]–[24].

### A. Models Accepting Raw Inputs

In terms of the format of input data to be processed, some modeling techniques rely directly on raw inputs [16], [19]–[23]. The work in [20] assumes a predefined topological knowledge about the object to be modeled while other solutions have been proposed in the framework of linear integer programming [19], [21], [22]. After building overcomplete connections between automatically identified nodes, optimal subgraphs are extracted by solving an optimization problem under specific constraints. These techniques are based on solving a theoretically non-deterministic polynomial-time hard (NP-hard) problem with increased complexity in case of scalable inputs. They also do not provide consistent modeling in different executions. In [23], a graph representation was generated by first initializing multiple short parametric curves along intensity ridges of the image. These open contours are

Canadian Institutes of Health Research (CIHR, 299166) operating grant and a Natural Sciences and Engineering Research Council of Canada (NSERC, 239876-2011) discovery grant to F. Lesage and NSERC Discovery Grant, RGPIN-2014-06089 to P. Pouliot.

[1] Institute of Biomedical Engineering, École Polytechnique de Montréal, Montreal, QC, Canada; [2] Department of Electrical Engineering, École Polytechnique de Montréal, Montreal, QC, Canada; [3] Research Centre, Montreal Hearth Institute, Montreal, QC, Canada; [4] Department of Computer and Software Engineering, École Polytechnique de Montréal, Montreal, QC, Canada. rafat.damseh@polymtl.ca

iteratively deformed by internal stretching forces and external image forces until saturation. Geometrical post-processing is then performed to connect the deformed contours and build a final graphical model. In [16], after calculating a Riemann vesselness potential, based on convolving the image with a steerable Laplacian of Gaussian filter banks, exhaustive search for geodesic connecting paths is performed to construct an over-connected graph. Then, a final minimum spanning tree is calculated. Methods accepting raw inputs are susceptible to other elongated structures in angiographic datasets and cannot ensure an alignment of the extracted graphs with vessels centerlines. Also, they require extra work to delineate the regions of interest and to eliminate unwanted compartments in the output vascular models. Finally, they can introduce over- or reduced-connectivity patterns, thus resulting in false topological interpretations.

### B. Models Accepting Voxel-based Mappings

Graphing solutions have also been proposed requiring pre-mapping of the object structure to be modeled [9]–[13], [24], [26]. Various techniques are surveyed in [27]. Here, we discuss those relying on voxel-based binar pre-mapping. Methods in [9], [10], [26] produce image-based curve skeletons following homotopic thinning. In [9], topological models for blood vessels are generated from 3D power doppler ultrasound images to differentiate changes in benign and malignant tumors. In [26], improved thinning performance is achieved by first shrinking the input volumetric binary mask through an iterative least squares optimization. Other image-based skeletonization techniques are based on calculating the singularities of distance-related fields [11]. Hybrid distance-based skeletonization is employed to provide a 3D path planning for virtual bronchoscopy in multidetector computed tomography. Thinning and distance-based skeletonization techniques are fast and easy to implement, however, they produce less-smooth skeletons with disconnected-compartments, that need extensive pruning and re-graphing.

### C. Models Accepting Geometry-based Mappings

The pre-mapping of vascular structures can also be a triangulated surface model or a point cloud. Geometric mapping of objects structure can yield graphical models with improved quality [12], [13], [24]. Methods to extract curve skeletons from incomplete point clouds are proposed in [13], [28]. In [28], the authors proposed an improved cloud-based graphing scheme exploiting piecewise cylindrical segmentations of the vascular structure. As reported in the corresponding papers, cloud-based schemes suffer from incorrect centeredness of the obtained models, especially in the case of irregular/non-sufficient sampling of the object structure. In [12], a mesh contraction scheme is introduced using differential geometry to decimate a surface model into a curve skeleton. In [24], mesh contraction is combined with a segmentation network and a surface mesh generator to model cerebral microvessels from raw TPM. Geometry-based skeletonization techniques require exhaustive efforts in preparing suitable surface model inputs, which is a major obstacle when working with complex and scalable cerebrovascular structures.

### D. Our Contribution

There is a need for vascular graphing schemes that are less restrictive to either hardly-encoded inputs or to high-quality vascular labeling while providing precise topological and structural representations to be used in further vascular analysis. Here, to address this issue, we propose a novel vascular graphing scheme inspired by the Laplacian flow formulation. Diverging from the work in [24], we exploit the Laplacian framework to deform 3D geometric graphs, instead of triangulated meshes, converting them into curve skeletons as models of vascular structures. Starting with a binary-delineated vascular structure, truncated 3D grid graphs are first constructed within vessels boundary. We develop a technique to assign affinity weightings to these graphs based on both the binary distance transform and the local geometry of graph compartments. The weighted graphs are fed into a constrained iterative optimizer to create a Laplacian dynamic flow of graph vertices/nodes toward the centerlines of vascular structures combined with a convergence criterion to stop the iteration process. Finally, a refinement algorithm is proposed to convert the deformed graph into a final vascular graphed-skeleton model. The proposed scheme can provide smooth and well-connected graph models regardless of the quality of image vascular pre-mapping, and yet does not require complex input representations, such as water-tight surface meshes. Our modeling is shown to provide improved vessel radii estimates, which are performed intrinsically during the geometric deformation process. This property is vital for biophysical studies that heavily rely on structural information encoded in vascular models. Furthermore, with simple hyper parameterization, the proposed scheme provides control over the calculation speed and the smoothness and centeredness properties of the generated graphs. The proposed scheme can be extended to process scalable vascular images at reasonable computational effort.

In the following, we present the formulation of the modeling scheme. We describe its parameter sensitivity and study performance on synthetic and real datasets acquired using different imaging modalities. Results are compared with that produced by other efficient and recent state-of-the-art graphing techniques. A brief conclusion follows.

## III. METHOD

In this section, we describe a novel computational geometry scheme to generate graph-based anatomical models of cerebrovascular networks based on their binary maps.

### A. Initial Geometry

We first construct an initial geometric graph enclosed within the boundary of the masked microvascular structure. We then aim at building a reduced grid graph with its nodes emerging from the true-valued voxels of the vascular mask. For $\mathcal{O} \subset P \subset \mathbb{R}^3$ being the object that represents the microvasculature in the image domain $P$, the corresponding 3D binary mapping is defined as $\mathcal{I} : P \to \{1, 0\} \mid \mathcal{I}(\mathbf{p}) = 1 \ \forall \mathbf{p} \in \mathcal{O}$ and $\mathcal{I}(\mathbf{p}) = 0$ otherwise. Furthermore, we define $\mathcal{D} : P \to \mathbb{R}$ to be the Euclidean distance transform obtained for $\mathcal{I}$. Now, one could define a reduced grid graph $\mathcal{G} = (\mathbf{V}, \mathbf{E})$, with $\mathbf{V} : \{1, 2, 3, \ldots, k\}$, where $k$ is equal to the number of voxels in $\mathcal{O}$. The labeling function $\phi_p$ assigns voxel coordinates $\mathbf{P} = [\mathbf{p}_1, \mathbf{p}_2, \mathbf{p}_3, \ldots, \mathbf{p}_k]^T, \mathbf{p}_i = [p_{xi}, p_{yi}, p_{zi}] \in \mathbb{R}^3$, to $\mathbf{V}$. Likewise, the function $\phi_r : \mathbf{V} \rightarrow \mathbf{R}$ maps graph nodes to $\mathbf{R} = [r_1, r_2, r_3, \ldots, r_k]^T, r_i = \mathcal{D}(\mathbf{p}_i)$. We construct the edges $\mathbf{E} : \{\{i,j\}, \forall i \in \mathbf{V} \text{ and } j \in \mathcal{N}(i)\}$, where $\mathcal{N}(i)$ is the set of nodes emerging from the voxels forming the 6-connectivity neighbors of the voxel associated with the node $i$. For an arbitrary voxel in 3D space, its 6-connectivity neighbors are the voxels sharing one of their surfaces with it. Employing other connectivity patterns for the construction of $\mathcal{G}$ is achievable, however, the 6-neighborhood pattern utilized in our scheme ensures a consistent performance at a lower computational burden.

### B. Geometric Flow and Graph Contraction

We follow the paradigm of Laplacian optimization to derive a geometric flow for the nodes of $\mathcal{G}$. One crucial characteristic sought for the dynamics of this flow is the inward contraction of $\mathbf{V}$. Precisely, the flow should ensure a convergence of nodes positions toward the centerline of the object $\mathcal{O}$. Now, let us consider $G$ as a weighted graph with affinity matrix $\mathbf{W}$. we define $\delta_{i,\mathcal{W}}$ as the laplacian operator applied on the $i$th node of $\mathcal{G}$ and acting on the labeling function $\phi_p$:

$$\delta_{i,\mathbf{W}} = \left[\sum_{j \forall \{i,j\} \in \mathbf{E}} \mathbf{W}^{ij} \phi_p(j)\right] - \phi_p(i) \ , \quad (1)$$

where $\sum_{j \forall \{i,j\} \in \mathbf{E}} \mathbf{W}^{ij} = 1$. To establish a geometric flow dynamics bearing the features discussed above, we propose a combined use of two different normalized affinity weightings on $\mathcal{G}$ defined as

$$\mathbf{W}_n^{ij} = \frac{\|\mathbf{p}_i - \mathbf{p}_j\|_2}{\sum_{k \forall \{i,k\} \in \mathbf{E}} \|\mathbf{p}_i - \mathbf{p}_k\|_2} \quad (2)$$

$$\mathbf{W}_m^{ij} = \frac{|r_i - r_j|}{\sum_{k \forall \{i,k\} \in \mathbf{E}} |r_i - r_j|} \ . \quad (3)$$

Henceforth, we will refer to $\mathbf{W}_n^{ij}$ and $\mathbf{W}_m^{ij}$ as $l$2-norm and medial affinity matrices, respectively, and the associated $\delta_{\mathbf{W}_n}$ and $\delta_{\mathbf{W}_m}$ as $l$2-norm and medial Laplacians. Based on a normalized affinity matrix $\mathbf{W}$ for $\mathcal{G}$, one can compute the Laplacians for the entire graph using the Laplacian matrix, i.e.,

$$\mathbf{L}_{ij} = \begin{cases} -1 & i = j \\ \mathbf{W}^{ij} & \{i,j\} \in \mathbf{E} \\ 0 & otherwise \ . \end{cases} \quad (4)$$

We denote the Laplacian matrices associated with the $l$2-norm and medial weights as $\mathbf{L}^n$ and $\mathbf{L}^m$, respectively. Consequently, the $l$2-norm and medial Laplacians of the entire graph are given by

$$\mathbf{\Delta}_n = [\delta_{1,\mathbf{W}_n}, \delta_{2,\mathbf{W}_n}, \ldots \delta_{k,\mathbf{W}_n}]^T = \mathbf{L}_n \mathbf{P} \ , \quad (5)$$

$$\mathbf{\Delta}_m = [\delta_{1,\mathbf{W}_m}, \delta_{2,\mathbf{W}_m}, \ldots \delta_{k,\mathbf{W}_m}]^T = \mathbf{L}_m \mathbf{P} \ . \quad (6)$$

Interestingly, in the formulas above, each of the elements $\delta_{k,\mathbf{W}_n}$ and $\delta_{k,\mathbf{W}_m}$, corresponding to a node $k$ in the graph, represents a local vector pointing toward a non-uniformly weighted centroid of its level-1 neighbors. The vector fields, $\mathbf{\Delta}_n$ and $\mathbf{\Delta}_m$, calculated based on a geometric grid graph generated from a labeled angiogram, are depicted in Fig. 1. Both fields comprise vectors that have different magnitudes and directed toward the centerline of the vessel tree. The field $\mathbf{\Delta}_n$ is zero within vessels boundary due to the equal spacing of grid graph nodes. Based on (1) and (2), this results into a zero net-vector. On the other hand, $\mathbf{\Delta}_m$ has non-zero magnitude even inside vessels body, since its calculated based on the Euclidean distance transform mapping.

Now, if graph nodes are further moved inward, both $\mathbf{\Delta}_n$ and $\mathbf{\Delta}_m$ will have smaller magnitudes in the direction perpendicular to the centerline. Hence, contracting graph geometry toward vessels centerlines can be obtained by implicitly solving

$$\begin{bmatrix} \mathbf{L}_n \\ \mathbf{L}_m \end{bmatrix} \widehat{\mathbf{P}} = 0 \ , \quad (7)$$

where $\widehat{\mathbf{P}}$ are the new positions of graph nodes. Since both matrices $\mathbf{L}_n$ and $\mathbf{L}_m$ are singular, this sparse system admits the trivial solution $\widehat{\mathbf{P}} = 0$. To address this issue, a regularized version of the system can be solved as

$$\begin{bmatrix} \mathbf{W}_\alpha \mathbf{L}_n \\ \mathbf{W}_\beta \mathbf{L}_m \\ \mathbf{W}_\gamma \end{bmatrix} \widehat{\mathbf{P}} = \begin{bmatrix} \mathbf{0} \\ \mathbf{W}_\gamma \mathbf{P} \end{bmatrix} \ . \quad (8)$$

Above, $\mathbf{W}_\alpha$, $\mathbf{W}_\beta$ and $\mathbf{W}_\gamma$ are diagonal matrices such that $\mathbf{W}_\alpha^i = \alpha$, $\mathbf{W}_\beta^i = \beta$ and $\mathbf{W}_\gamma^i = \gamma$, respectively. The elements in $\mathbf{W}_\gamma^i = \gamma$ constrain all graph nodes to their current positions. Those in $\mathbf{W}_\alpha$, $\mathbf{W}_\beta$ determine the dependence of the solution on either the 2-norm or the medial Laplacians, respectively.

### C. Contraction Flow Dynamics

Solving (8) iteratively generates a geometric flow for graph nodes by moving them along the direction of their 2-norm or the medial Laplacians. This flow dynamics will serve as the core of the proposed skeletonization scheme. The nature of the flow is controlled by varying the weights $\alpha$, $\beta$ and $\gamma$. Several iterations, and proper weights, are required for the process to converge towards the centerline of vascular structures. The proposed model creates a dual-mechanism to guide the dynamics of graph nodes by exploiting both, their connectivity-encoded geometry and distance transform mapping. This duality is vital to ensure a better convergence of graph geometry. Using only the weights from $\mathbf{L}_n$ could lead to unwanted straightening of the graph geometry, and integrating $\mathbf{L}_m$ weights in the dynamics will restrict flow dynamics toward singularities/ridges emerging in the distance transform field. On the other hand, relying only on distance field weightings does not guarantee a convergence into a curve skeleton with centeredness property [29]. Despite a less-complex computation, the distance transform has the shortcoming of forming medial surfaces rather than medial curves in the case of 3D objects. Thus, combining $\mathbf{L}_n$ weights in (8) provides additional constraints on the flow dynamics to impose further shrinking of graph geometry. At each iteration, the solution of the over-determined system in (8) is obtained by least-squares minimization. The parameters $\alpha \in [0, 1]$ and $\beta \in [0, 1]$ were chosen to have a partition of unity property, and thus only one of the two needs to be specified. The speed of flow is thus tuned by altering the value of $\gamma \in [0, 1]$. Lower values of $\gamma$ accelerate the contraction process at the expense of over-deformed graph geometry. Conversely, higher values of $\gamma$ decelerate the geometric evolution of graph nodes but can result in slow convergence. The final concern is to define a criterion to stop the contraction process when converged. For this purpose, at an iteration $t$, we calculate the areas of polygons derived from the cycle basis of the graph [30]. Then, we retain the cumulative sum of areas, denoted as $A_t$, for polygons containing a number of nodes less than $\epsilon_n$. The iteration stops when the ratio of $A_t$ and $A_0$ is smaller than a threshold, referred to as $\epsilon_A$. In the following, the values of $\epsilon_n$ and $\epsilon_A$ were set empirically to 10 and $10^{-3}$, respectively.

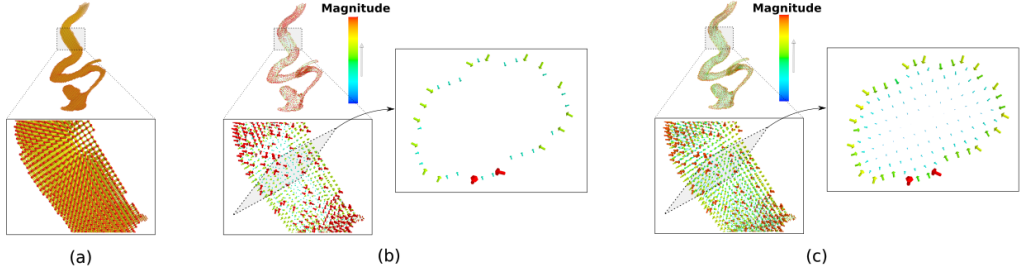

Fig. 1: Visualization of the $\mathbf{\Delta}_n$ and $\mathbf{\Delta}_m$ vector fields, depicted in (b) and (c), and calculated using (5) and (6), respectively. The two fields result from applying the proposed Laplacian operators on a geometric graph enclosed within vascular space, shown in (a). Vectors colors and scales reflect their magnitudes, which are bounded between 0 and 1.

### D. Adaptive Flow Dynamics

During the contraction process, some nodes in the graph could evolve towards skeleton-like branch node characteristics. Here, we propose a technique to alleviate the dynamics of these nodes in following time steps. The technique calculates the angles that are formed between any pair of edges arising from a node $i$. If the absolute cosine of these angles is close to one, that node is more-likely to be a branch node. For a node $i$ to undergo a reduced contraction force in the next time step, it has to satisfy the condition:

$$|\cos(\pi - \angle [\overrightarrow{\mathbf{p}_i\ \mathbf{a}}, \overrightarrow{\mathbf{p}_i\ \mathbf{b}}])|, \ \mathbf{a}, \mathbf{b} \in \mathcal{N}(\mathbf{p}_i), \ \geq 0.9 \ . \quad (9)$$

When a node $i$ passes the above condition, its corresponding element in $\mathbf{W}_\gamma$ is altered to $\mathbf{W}_\gamma^i = \epsilon_\gamma * \gamma$. The value of the factor $\epsilon_\gamma$ should be chosen carefully. Larger values of $\epsilon_\gamma$ can disrupt the row/column scaling in the system, thus hindering its convergence. On the other hand, smaller values are not helpful in restraining skeletal-like nodes. Based on experimentations, setting $\epsilon_\gamma$ to 10 was a good choice for the modeling done below.

### E. Geometric Graph Decimation

Following contraction, graph nodes are squeezed after each iteration forming a denser geometry while shifting closer to the object centerline in each step. Yet, the topology of the graph, remains the same. To address this, we take advantage of the collapsed geometry to adjust graph structure. The structural surgery includes a reduction of the graph and a modification of its topology after each iteration. This decimation process helps in deforming the graph to approach that of a curve skeleton. Also, it delivers a reduced model that will require less computational effort when processed in the next iteration. Consequently, the overall contraction time will be minimized. First, the Euclidean space is tessellated into a 3D regular grid with equally-spaced cubic cells of size $c^3$. Practically, we used a cell size equal to the size of a voxel in the 3D input image. Graph nodes situated within a cell are then grouped as one cluster. After generating the clusters, $\mathcal{C} : \{\mathbf{C}_1, \mathbf{C}_2, \mathbf{C}_3, \ldots, \mathbf{C}_k\}$, their centroids, $\mathbf{S} : \{\mathbf{s}_1, \mathbf{s}_2, \mathbf{s}_3, \ldots, \mathbf{s}_k\}$, are calculated and integrated as new nodes in the graph. Then, new edges are assigned for the newly created nodes and an edge between a pair of centroid nodes is created if there is any connection between their corresponding samples. Finally, we remove graph nodes used in clustering, and only the new ones, with their constructed edges, are maintained. The structural surgery described above is depicted in figure 2.

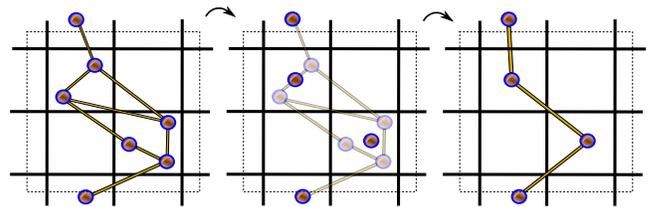

Fig. 2: An illustration of the proposed clustering technique used for decimation of the evolving geometric graph at each iteration.

### F. Transferable Radius Attributes

Values at the ridges formed by the distance transform provides a good estimation of the radii of tubular objects along the trace of their centerlines. As previously discussed, a graph node $i$ has the attribute $r_i$ representing the closest Euclidean distance of the node from the object boundary. If the node is located at the medial curve of the object, the value of $r_i$ can serve as a measure of the object radius at that point. Based on the proposed contraction scheme, existing graph nodes are replaced at each time step with those newly created through our structural operation. When applying the structural surgery, a new node $i$, emerging as the centroid of the cluster $\mathbf{C}_i$, will be assigned an attribute $r_i$ computed as the maximum of those associated with the nodes in $\mathbf{C}_i$. Using this technique, the graphed skeleton will not only model objects topology, but also their shapes, an essential input to cerebrovascular models in wider biophysical applications.

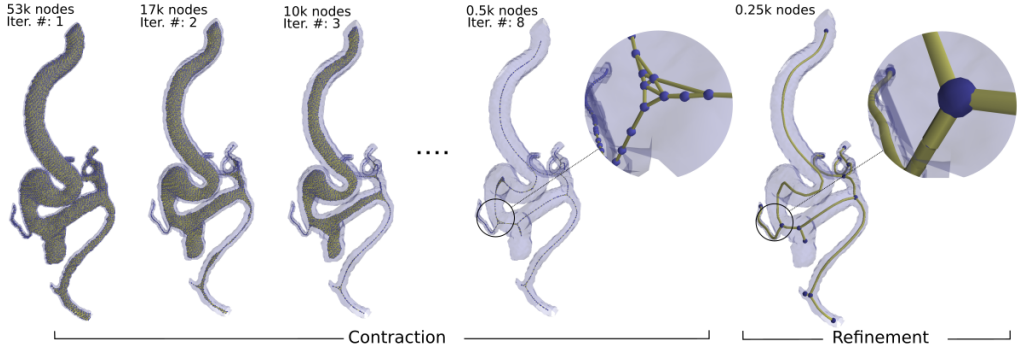

Fig. 3: Visual example of the contraction and refinement processes integrated in the proposed scheme to produce a final graphed-skeleton from a binary-labeled vascular structure. In the contraction phase from left to right, the initial geometric graph evolves toward vessels centerline with further structural reduction at each iteration. The refinement phase accounts for fixing over-connectivity patterns on the contacted graph, converting it into a graphed curve skeleton.

An example that illustrates the proposed contraction process is depicted in Figure 3 (contraction phase). It shows how a decimated version of the graph $\mathcal{G}$ is moved towards the object medial line at each iteration. The algorithmic pipeline for the contraction scheme is summarized in Algorithm 1.

---

**Algorithm 1** Structural processing of $\mathcal{G}$ in the contraction phase.

---

**Require:** $\mathcal{G}, \alpha, \beta, \gamma$
1: Compute $A_0$; $A_t \leftarrow A_0$.
2: **while** $\frac{A_t}{A_0} < 10^{-3}$ **do**
3:  Compute $\mathbf{L}_n$, $\mathbf{L}_m$ based on (2), (3) and (4). (Section III-B)
4:  Obtain $\mathbf{W}_\gamma$ after imposing each $\mathbf{p}_i \in \mathbf{P}$ to (9). (Section III-D)
5:  Solve (8). (Section III-B)
6:  Create $\mathcal{C}$ and apply graph decimation. (Section III-E)
7:  Calculate radii attributes. (Section III-F)
8:  Update $A_t$. (Section III-C)
9: **end while**

---

### G. Geometric Graph Refinement

The output graph resulting from the contraction scheme is converged to the centerline of the object $\mathcal{O}$ but does not have the skeleton structure properties required for modeling. In this section, we propose a refinement operation to convert the geometric graph $\mathcal{G}$ resulting from Algorithm 1 into a curve-skeleton. The procedure is based on eliminating small polygons formed in $\mathcal{G}$. Based on experimentation's, a polygon is considered small if its area is less than $10c^2$. The cycle basis of $\mathcal{G}$ is computed and the set of cycles $\mathcal{X}$ with edges $\mathbf{E}_{\mathbf{X}_i}$ in the $i$th cycle, is retained as follows:

$$\mathcal{X} = \left\{ \mathbf{X}_i, \text{ if } \frac{1}{2} \left\| \sum_{(\mathbf{a},\mathbf{b}) \in \mathbf{E}_{\mathbf{X}_i}} [\overrightarrow{O\mathbf{a}} \times \overrightarrow{O\mathbf{b}}] \right\| < 50 \right\}, \quad (10)$$

where $O$ is the origin. Nodes in the cycle $\mathbf{X}_i$ are contracted towards their centroid $\mathbf{s}_i$:

$$\mathbf{p} \leftarrow (\mathbf{p} + \mathbf{s}_i)/2, \quad \forall \mathbf{p} \in \mathbf{X}_i. \quad (11)$$

If a node belongs to multiple cycles, it is randomly processed within one of these cycle. After adjusting nodes geometry, we carry out the same structural surgery presented in the contraction stage. We repeat the steps discussed above until we reach a state where the number of elements in $\mathcal{X}$ is equal to zero. Figure 3 visualizes how the proposed refinement technique transforms a decimated graph resulting from the contraction process into a graph-based skeleton.

## IV. VALIDATION AND DISCUSSION

In this section, we evaluate the performance of the proposed modeling scheme through various experiments conducted on synthetic and real cerebrovascular datasets. The data includes both tree-like and loopy vascular structures. Fist, we present the evaluation metrics, the different datasets and baselines utilized for performance evaluation. We study the sensitivity of our modeling to its tunable parameters. Finally, We compare our results with that obtained by other standard and state-of-the-art skeletonization approaches [12], [16], [23], [31], [32] when applied to the same datasets.

### A. Experimental Setup and Datasets

We compared the performance of our modeling scheme with other efficient methods used to generate graphed skeletons for tubular or vessel-like structures. These methods are: the classical 3D-thinning (Voxel-based) [9][1], mesh-based mean-curvature skeletonization (Mesh-based) [12][2], 3D points-cloud contraction (PC-based) [13][3], multiple stretching of open active contours (SOAX-based) [23][4] and geodesic minimum spanning trees extraction (VTrails-based) [16][5]. For quantitative assessments of the structural errors produced by the various skeletonization methods, we use the DIADEM metric [33][6]. This metric provides a value $\in [0, 1]$, with 1 indicating a perfect model match. It is to be noted the DIADEM metric is used for evaluating only tree vascular structures and is not applicable for acyclic vascular networks. The NetMets measures [34] are also exploited to assess the quality of both tree-like and loopy vascular structures. These measures are employed to provide detailed quantification of the geometrical and topological errors incurred by the different schemes. A code for generating these measures is available online [7]. The associated measures contain four metrics. Two metrics, namely GFNR and GFPR, are used to evaluate the geometric false negative and false positive errors, respectively. The other two metrics, namely, CFNR and CFPR, quantify the false negative and false positive topological errors, respectively. The sensitivity of NetMets measures is controlled by one parameter denoted by $\tau$. For more details on these metrics, the reader is referred to [34].

---
[1]Implementation: *https://github.com/InsightSoftwareConsortium/ ITKThickness3D*
[2]Implementation: *https://github.com/CGAL/cgal*

We based our evaluation on both synthetic and real cerebrovascular structures captured in in-vivo studies using different acquisition modalities. We used the VascSynth toolbox [35] to prepare 3 sets of synthetic vascular trees, with 30 samples in each, $D^{nl}$, considering different levels of surface noise $nl = low, medium$ and $high$. Each set contains samples with three different branching levels (10 samples for each level), namely, $16, 32$ and $64$. Surface noise was created by adding bubbles on vessel boundaries. The higher the noise level, the higher is the number and size of these bubbles. The clinical validation set is composed of x-ray cerebral aneurysm angiographies (RAA) [36] and magnetic resonance angiographies (MRA) [37]. We also assessed the algorithm on in-vivo two-photon microscopy stacks (TPM) obtained from [38]. Ground truth annotations from the various datasets are provided as spatial centerlines. To extract graph models, binary maps of vascular structures were fed to the schemes that accept image-based inputs. This procedure is followed even for the schemes that allow raw inputs to ensure a fair comparison. For the mesh-based and PC-based methods, since they do not accept voxelised datasets, we supplied surface models generated from the 3D binary angiograms using the algorithm described in [24]. The extracted meshes are processed directly by the mesh-based skeletonizer whereas its vertices are used as inputs to the PC-based method. After generating an output graph using any of the schemes, radii values are assigned to its nodes. Vascular radius at each graph node is quantified as the intensity of the Euclidean distance transform at the coordinate of that node. We assess the correctness of radius mappings on the generated graph by measuring the mean absolute percentage error (MAP) between the estimated and true radii values. To compute MAP scores, we first match the nodes of the experimental model with that of the ground truth as explained in [24]. All of our experiments were executed on a 3.0 GHz Ryzen AMD processor (8 cores, 16 threads in each) with 64 GB of RAM.

### B. Parameters Sensitivity

The parameters that needs to be tuned in our methodology are $\alpha$, $\beta$ and $\gamma$. This section provides insights on the behavior of the algorithm when varying these parameters. Synthetic objects, depicted in Figure 4, were skeletonized at different values of $\alpha$, $\beta$ and $\gamma$. The upper row displays a tubular object with large boundary perturbations, i.e., less-informative distance transform mapping about the tubular structure. In the upper row, after solving (8) for 10 iterations, it is clearly seen that enforcing the contraction process to rely only on medial Laplacians, $\mathbf{L}_m$ hinders the convergence toward object centerline. The deformed model remains on object medial surfaces. On the other hand, when increasing the effect of the $l2$-norm Laplacians by increasing the value of $\alpha$, better convergence and centerline modeling is achieved. In the middle row of the figure, a bent tubular structure inflated at its center is used in the modeling process. When cancelling medial Laplacians $\mathbf{L}_m$ by setting $\beta$ to zero, the output graph is abstract and fails in capturing the full information about the object. Nevertheless, adding $\mathbf{L}_m$ in the optimization problem creates better modeling. Here, we conclude by stating that solving (8) with a combined setting of both $\alpha$ and $\beta$ can provide smooth, and yet, well-detailed graph models. As previously mentioned, the speed of the contraction process can be controlled by $\gamma$. In the bottom row of the figure, from left to right, an 8-shape object is modeled with decreasing values of $\gamma$, namely, $0.1, 0.05, 0.005, 0.001$. The number of iterations required to converge were $34, 17, 9$ and $5$, respectively. It is observed that the lower the value of $\gamma$, fewer iterations are needed, but at the expense of more abstract graphical representation. It is also seen that abstracted graphs suffer from poor alignment of their compartments along the object centerline. Based on this empirical assessment, in studies of vascular structures acquired with different imaging modalities, we used $\alpha = 0.5$, $\beta = 0.5$ and $\gamma = 0.05$ as the parametric setup.

### C. Synthetic Vascular Trees

Table I lists the error measures based on the metrics used for each graphing scheme on synthetic angiograms. For each metric, the top-three scores are color-labeled, while the top-score is green-labeled. The arrow next to each metric name indicates if a high (up-arrow) or low (down-arrow) score is better. When assessing the extracted geometry based on GFNR and GFPR metrics, one can note that at low level of boundary noise, the voxel-based method provides the lowest false negative and positive rates. However, with increased noise level, it fails by introducing large geometric errors. The voxel-based method exhibits variable robustness against different levels of boundary noise. The SOAX-based method provides low false positive errors but at the expense of very high false negative rates. This observation indicates that it misses substantial geometric details of the vascular structures. Overall, our algorithm performs better against other schemes in providing more robust graph models at high noise levels, with, overall, the lowest false negative and false positive rates. Analysing the topological CFNR and CFPR errors, the voxel-based method gives low false negative scores but at the expense of a degradation in false positive rates. This reflects a deterioration of the original topology with a large amount of over-connectivity. The Vtrails-based method provides balanced, yet weak false negative and positive rates. The mesh-based technique shows a similar trend with some improvements. When considering both CFNR and CFPR scores, the proposed work stands again as the best scheme to provide the best error-less topological representations. We then analysed the tree-like correctness of the graphs generated by the various schemes based on the DIADEM metric. The mesh based scheme provides comparable results to ours in some cases. Yet, the proposed method shows superior performance over all other methods in most of the cases. Finally the MAP metric is of vital importance for validating the various techniques at providing comprehensive anatomical models of the vascular structure. It is very clear from the table that, in all cases, the method described here was the best at capturing of anatomical information.

---

[3] Implementation: *http://www.shihaowu.net*
[4] Code: *https://github.com/tix209/SOAX*
[5] *https://github.com/VTrails/VTrailsToolkit*
[6] *http://diademchallenge.org/metric.html*
[7] *https://git.stim.ee.uh.edu/segmentation/netmets*

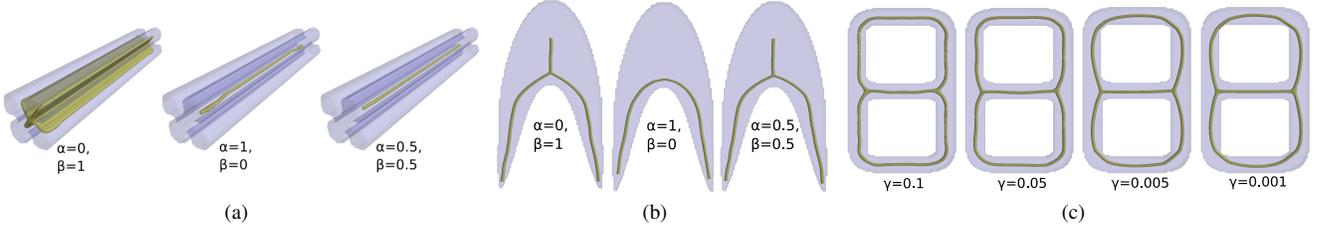

Fig. 4: The response of the proposed scheme when modeling various structures using different parametric settings. In (a) and (b), the skeletonization output is robust to variations in the input structure when using a combined setting of $\alpha$ and $\beta$. Varying the value of $\gamma$ impacts the quality of the output graph as seen in (c).

TABLE I: Error scores calculated after applying the different graphing schemes on the synthetic vascular dataset. Colored cells indicate the best 3 scores while a green cell indicates the best score.

| | GFNR ↓ | | | | | | | GFPR ↓ | | | | | |
|---|---|---|---|---|---|---|---|---|---|---|---|---|---|
| | Voxel-based | Mesh-based | PC-based | SOAX-based | Vtrails-based | Proposed | | Voxel-based | Mesh-based | PC-based | SOAX-based | Vtrails-based | Proposed |
| $D^0$ | **0.15±0.01** | 0.27±0.03 | 0.23±0.04 | 0.35±0.05 | 0.32±0.03 | 0.22±0.03 | $D^0$ | **0.11±0.02** | 0.14±0.02 | 0.33±0.04 | 0.15±0.03 | 0.20±0.03 | 0.12±0.02 |
| $D^1$ | 0.27±0.03 | 0.28±0.04 | 0.32±0.07 | 0.40±0.05 | 0.37±0.05 | **0.20±0.02** | $D^1$ | 0.28±0.05 | 0.19±0.04 | 0.31±0.07 | 0.16±0.02 | 0.23±0.05 | **0.16±0.03** |
| $D^2$ | 0.41±0.09 | 0.34±0.04 | 0.35±0.09 | 0.44±0.06 | 0.42±0.06 | **0.26±0.03** | $D^2$ | 0.33±0.06 | 0.24±0.04 | 0.25±0.04 | 0.19±0.03 | 0.23±0.05 | **0.17±0.03** |
| | CFNR ↓ | | | | | | | CFPR ↓ | | | | | |
| | Voxel-based | Mesh-based | PC-based | SOAX-based | Vtrails-based | Proposed | | Voxel-based | Mesh-based | PC-based | SOAX-based | Vtrails-based | Proposed |
| $D^0$ | 0.27±0.05 | 0.37±0.04 | 0.49±0.11 | 0.56±0.09 | 0.52±0.06 | **0.23±0.06** | $D^0$ | **0.07±0.03** | 0.12±0.04 | 0.25±0.05 | 0.13±0.05 | 0.11±0.03 | 0.08±0.02 |
| $D^1$ | **0.24±0.05** | 0.41±0.11 | 0.54±0.08 | 0.62±0.16 | 0.57±0.09 | 0.28±0.07 | $D^1$ | 0.27±0.06 | 0.17±0.07 | 0.31±0.05 | 0.13±0.04 | 0.19±0.08 | **0.11±0.04** |
| $D^2$ | **0.17±0.03** | 0.34±0.07 | 0.51±0.11 | 0.67±0.14 | 0.53±0.13 | 0.26±0.05 | $D^2$ | 0.57±0.11 | 0.32±0.06 | 0.37±0.08 | 0.16±0.04 | 0.21±0.05 | **0.10±0.03** |
| | DIADEM(%) ↑ | | | | | | | MAP(%) ↓ | | | | | |
| | Voxel-based | Mesh-based | PC-based | SOAX-based | Vtrails-based | Proposed | | Voxel-based | Mesh-based | PC-based | SOAX-based | Vtrails-based | Proposed |
| $D^0$ | 65.1±8.22 | 72.54±9.8 | 63.2±12.1 | 66.9±6.44 | 58.4±8.82 | **74.8±7.29** | $D^0$ | 13.4±0.84 | 14.5±1.44 | 21.7±2.13 | 18.2±1.59 | 18.4±1.46 | **11.3±1.67** |
| $D^1$ | 54.1±13.5 | 69.3±8.2 | 51.1±14.0 | 56.5±10.4 | 55.9±7.35 | **71.7±6.42** | $D^1$ | 14.1±1.34 | 16.8±2.13 | 19.2±1.97 | 18.4±2.07 | 20.8±3.07 | **12.1±1.87** |
| $D^2$ | 42.4±7.06 | 64.4±9.45 | 44.5±13.8 | 49.3±9.91 | 56.2±9.3 | **67.1±8.15** | $D^2$ | 14.8±1.84 | 16.4±2.39 | 22.6±1.70 | 19.2±1.37 | 19.6±2.19 | **12.7±1.43** |

### D. Clinical Vascular Trees

| | DIADEM(%) ↑ | | | | | |
|---|---|---|---|---|---|---|
| | Voxel-based | Mesh-based | PC-based | SOAX-based | Vtrails-based | Proposed |
| RAA | 81.6±9.05 | 83.6±8.64 | 54.5±12.9 | 75.8±12.0 | 78.4±6.93 | **89.5±7.38** |
| MRA | 85.5±6.19 | 90.4±3.61 | 69.6±5.07 | 78.5±9.40 | 77.1±8.85 | **92.8±4.29** |
| | MAP(%) ↓ | | | | | |
| | Voxel-based | Mesh-based | PC-based | SOAX-based | Vtrails-based | Proposed |
| RAA | 8.08±1.88 | 8.22±1.81 | 18.9±5.28 | 10.6±1.75 | 11.7±2.60 | **6.03±1.73** |
| MRA | 11.5±3.88 | 10.8±2.39 | 14.9±0.24 | 15.1±0.97 | 15.8±1.36 | **4.77±0.23** |
| TPM | 7.73±2.36 | 8.93±3.44 | 14.1±4.65 | 10.5±3.58 | –n/a– | **6.37±2.35** |

TABLE II: DIADEM and MAP measurements obtained after applying the various graphing schemes on the RAA and MRA datasets. MAP metric is also used to assess the performance on the TPM dataset.

Figure 5 (a) and (b) depicts the NetMets measures when testing the graphing schemes on RAA and MRA datasets, respectively, at different tolerance levels $\tau$. Again, the proposed scheme was capable of providing superior models in all cases. SOAX-based and VTrails-based methods provide graph models with disparate topological false negative and false positive rates, hence indicating under- or over-connected graph models. This is confirmed qualitatively later in the discussion. The remaining schemes, including ours, are proven to deliver better geometrical/topological models. In most cases of RAA and MRA datasets, the proposed modeling scheme outperformed all other methods. We also studied the correctness of the tree-like structures obtained by running the minimum spanning tree algorithm on the graphs extracted.

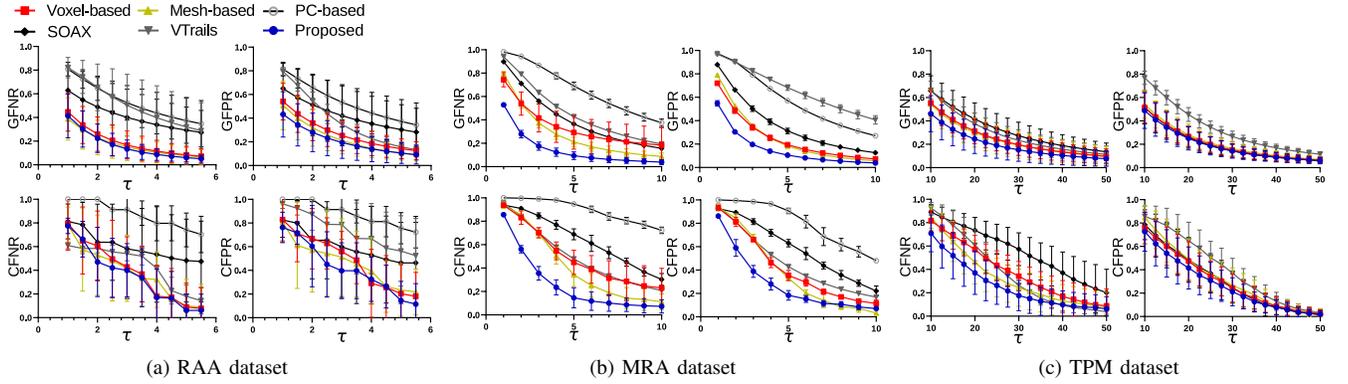

(a) RAA dataset  (b) MRA dataset  (c) TPM dataset

Fig. 5: NetMets measures obtained at different tolerance levels $\tau$ after applying the various schemes on real datasets.

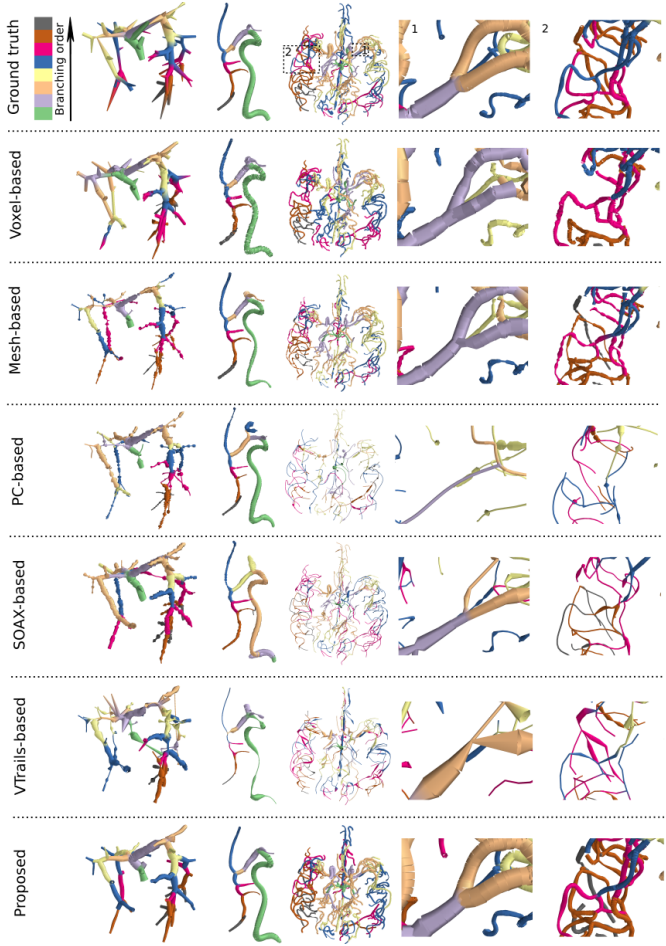

Fig. 6: 3D rendering of the vascular tree models obtained using the various schemes when applied to samples from synthetic (left column), RAA (second-left column) and MRA (middle column) datasets. The last two right columns depict magnified regions from the MRA models.

Arterial trees were constructed from MRA models based on the root branching from the Circle of Willis. The root arteries in RAA graphs were identified based on the vessel size and shape. Table II lists the DIADEM errors calculated from mapping the MRA and RAA trees to their corresponding ground truths. It is clear from the table that the proposed scheme performed substantially better in providing accurate minimum spanning trees of vascular structures with about 90% of ground truth similarity. The anatomical structure of the extracted trees were also assessed based on the MAP metric. Again, the proposed scheme outperformed other schemes in correctly mapping radius information, reflecting a more acurate anatomy. For visual assessment, examples of tree-like graphs, generated using the various schemes to model RAA and MRA angiograms, are depicted in Figure 6. The branching level along the tree is coded in color while radius information is mapped as the scale of cylinders (their diameters) capturing vessel segments. As shown in the figure, PC-based , SOAX-based and VTrails methods produce poorly connected graphs with degraded radius mapping and errors in their tree-like representations. The voxel-based method produce less-smooth graphs with many discontinuities. The mesh-based method is providing improved smoothness and inter-connectivity but with poor anatomical mapping. Nevertheless, in all cases, the proposed method is clearly shown to provide accurate, and yet smooth tree-like graphs with substantial improvement on the anatomical modeling aspect based on radius information.

### E. In-vivo Microvascular Networks

We then studied more complex microscopic angiographies acquired with TPM. It is to be noted that tree-like models cannot be extracted from these angiograms since they include loopy structures. Thus assessments are based on NetMets and MAP metrics. NetMets measures are depicted in Figure 5 (c). It is seen from the figure that the proposed method produces models having the lowest geometric and topological errors compared to all other methods. It is also seen that the PC-based and SOAX-based methods are not suitable for modeling such complex angiograms. It should be noted that the Vtrails-based method is not applicable in this case since it is specific to tree-like structures. The voxel-based method tends to create graphs missing many topological details, as seen in the CFNR plot.

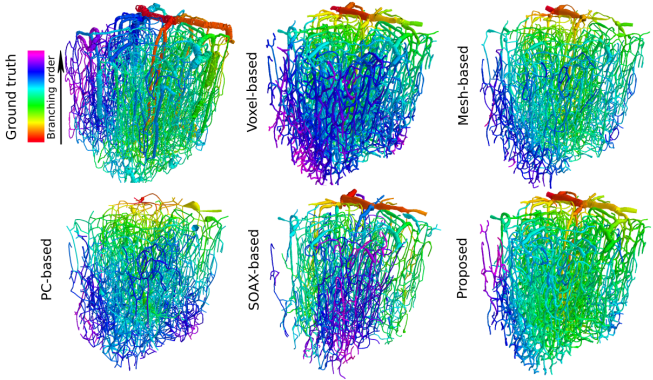

Fig. 7: Graph models generated for one of the TPM angiograms using the various skeletonization schemes.

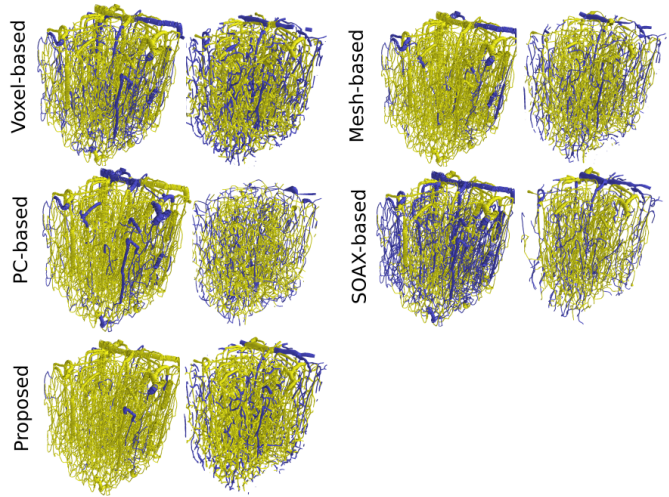

Fig. 8: For each method, we visualize the topological errors, namely, CFNR (left) and CFPR (right), measured on the models in Figure 7. Errors are blue-coded.

Mesh-based technique can provide acceptable graph models, however, its requirement of hardly-encoded surface models as inputs can hinder its applicability, especially in case of unclean TMP data. Figure 7 shows the 3D rendering of one TMP graph produced using the different methods. The figure depicts the propagation across the branches in the microvascular network by manually selecting a descending arterial as a source branch. These directed graphs are produced following a search from the source branch. Again, the figure illustrates the good performance of the Mesh-based and the proposed schemes in providing proper models compared to that of other schemes. Yet, our methodology is superior when considering the anatomical structures represented in vessels radii. Figure 8 visualizes the topological errors, namely, CFNR and CFPR, incurred in the graphs presented in Figure 7. Compared to all other methods, the proposed scheme is capable of generating TPM vascular models having less false negative and false positive topological errors. This coincides the results obtained in Figure 5 (c).

## V. COMPUTATIONAL TIME

For a comprehensive assessment of the various graphing schemes, we plot in Figure 9 their average computation time needed to process the angiograms in the RAA, MRA and TPM datasets. As seen from the figure, the Voxel-based technique scores the lowest computational effort. However, this speed improvement comes at the expense of unreliable topological results, that suffer from false negative connectivity patterns as previously shown from the CFNR measures in Figure 5 and disconnected components as shown in Figures 6. Comparing with other schemes, apart for the voxel-based method, the proposed scheme reduces the computational time by a large margin. Based on its improved modeling results at a lower demand of computational power, our scheme stands as a suitable alternative to the currently available graphing techniques for processing binary-labeled angiographic datasets.

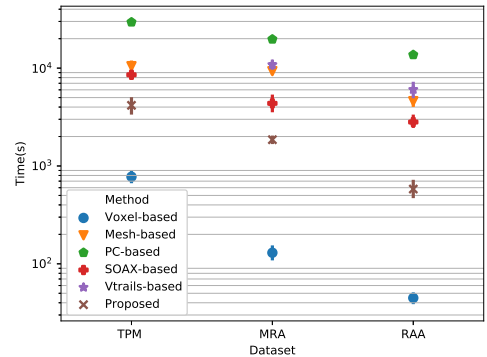

Fig. 9: Computation times (plotted as average and standard deviation) required by the various graphing schemes when applied to different datasets.

## VI. CONCLUSION

Overcoming current challenges in vascular graph-based modeling using available and state-of-the-art methods, which produce less-accurate connectivity patterns and under-complete anatomical features, a novel graphing approach has been proposed. The proposed scheme works on binary maps of vascular structures regardless of its quality, and is not restricted to hard-coded inputs like water-tight surface models. The proposed work is inspired by the Laplacian flow formulation used for mesh processing in the field of differential geometry. First, an initial geometric graph, in the form of a truncated 3D grid graph, is created filling the vascular space. A procedure for assigning affinity weights on the initial graph has been described. Based on these weights, we derived the Laplacian optimization problem to be solved iteratively, thus generating a dynamic evolution of the initial geometry toward vascular centerlines. We have designed a full algorithmic scheme to stop the iterative process when converged and apply a refinement surgery to convert the evolved geometry into a graphed skeleton. Our scheme integrates a local and intrinsic vessels radii calculations during the evolution and refinement stages. We validated the proposed scheme on synthetic and real angiographies and compared our results with those extracted by other efficient and state-of-the-art graphing schemes. The results support that the proposed algorithm provides more accurate vascular models holding better anatomical features.